\documentclass[%
reprint,
superscriptaddress,
 amsmath,amssymb,
 aps,
]{revtex4-2}

\usepackage{graphicx}
\usepackage{dcolumn}
\usepackage{bm}
\usepackage{color}

\begin{document}

\preprint{APS/123-QED}

\title{Magnesium Oxide at Extreme Temperatures and Pressures\\ Studied with First-Principles Simulations}

\author{Fran\c{c}ois Soubiran}
 \email{francois.soubiran@ens-lyon.org}
  \affiliation{Department of Earth and Planetary Science, University of California, Berkeley, CA 94720, USA}
\affiliation{\'Ecole Normale Sup\'erieure de Lyon, Universit\'e Lyon 1, Laboratoire de G\'eologie de Lyon, CNRS UMR 5276, 69364 Lyon Cedex 07, France}

\author{Felipe Gonz\'alez-Cataldo}
 \affiliation{Department of Earth and Planetary Science, University of California, Berkeley, CA 94720, USA}

\author{Kevin P. Driver}
 \affiliation{Department of Earth and Planetary Science, University of California, Berkeley, CA 94720, USA}
 \affiliation{Lawrence Livermore National Laboratory, Livermore, California 94550, USA}

\author{Shuai Zhang}
  \affiliation{Department of Earth and Planetary Science, University of California, Berkeley, CA 94720, USA}
 \affiliation{Laboratory for Laser Energetics, University of Rochester, Rochester, NY 14623, USA}
 \date{\today}

\author{Burkhard Militzer}
 \email{militzer@berkeley.edu}
\affiliation{Department of Earth and Planetary Science, University of California, Berkeley, CA 94720, USA}
 \affiliation{Department of Astronomy, University of California, Berkeley, CA 94720, USA}

\begin{abstract}
We combine two first-principles computer simulation techniques, path integral Monte-Carlo and density functional
theory molecular dynamics, to determine the equation of state of magnesium oxide in the regime of warm dense matter, with densities ranging from 0.35 to 71~g$\,$cm$^{-3}$ and temperatures from 10,000 K to $5\times10^8$~K. These conditions are relevant for the interiors of giant planets and stars as well as for shock wave compression measurements and inertial confinement fusion experiments. We study the electronic structure of MgO and the ionization mechanisms as a function of density and temperature. We show that the L-shell orbitals of magnesium and oxygen hybridize at high density. This results into a gradual ionization of the L-shell with increasing density and temperature. In this regard, MgO behaves differently from pure oxygen, which is reflected in the shape of the MgO principal shock Hugoniot curve. The curve of oxygen shows two compression maxima, while that of MgO shows only one. We predict a maximum compression ratio of 4.66 to occur for a temperature of 6.73 $\times 10^7$ K. Finally we study how multiple shocks and ramp waves can be used to cover a large range of densities and temperatures. 
\end{abstract}
\maketitle

\section{Introduction}
The discovery of numerous exoplanets~\cite{ExoplanetArchive} has
renewed the interest in materials at extreme conditions since their
behavior determines a planet's internal structure and
evolution. Laboratory experiments and computer
simulations have both contributed significantly to
the characterization of matter at extreme conditions. Developing
accurate equations of state (EOS) for candidate materials is necessary
not only to infer planetary formation scenarios, but also to
understand how materials respond to extreme conditions including those
present in high-velocity impacts, shock compression, and inertial
confinement fusion (ICF) experiments~\cite{Hammel2010}. These
applications require an EOS that is able to describe different degrees
of ionization, as the temperatures can increase by several orders of
magnitude during these dynamic compression events. Inconsistencies in
the EOS can lead to incorrect interpretations of the experiments or to
contradicting planet formation scenarios that arise, for example, from different
relations between isentropes and melting
curves~\cite{Fratanduono2018}.

Magnesium oxide (MgO) is one of the most abundant minerals in the
Earth's interior and is considered to be a primary building block of
planetary formation~\cite{Valencia2010,Bolis2016}. It is commonly used
as a representative mantle material in exoplanet
modeling~\cite{Seager2007,Valencia2009,Valencia2010,Wagner2012}.
Shock compression experiments on SiO$_2$~\cite{Hicks2006,Millot2015}
and MgO~\cite{Mcwilliams2012,Miyanishi2015,Bolis2016}, combined with
first principle calculations, demonstrated that these mantle minerals
become electrically conducting in the fluid phase. Super-Earth planets can thus generate magnetic fields within their
mantles~\cite{Soubiran2018}.  At ambient conditions, MgO has the
rock-salt (B1) structure which is known from experiments to be stable
up to at least 227 GPa~\cite{Duffy1995}. Around 500 GPa, it is
predicted to transform into the CsCl (B2) structure~\cite{Belonoshko1996,Boates2013,Bouchet2019}, which is believed to be the only stable phase up to at least 40 Mbar, a pressure at which MgO remains solid at least up to
20\,000 K~\cite{Wilson2012}. Together with silica (SiO$_2$), MgO is one of
the endmembers on the last-stage dissociation of
MgSiO$_3$~\cite{Boates2013,Tsuchiya2004,Umemoto2011,Umemoto2017}.

Recent experiments using plate impacts have measured the shock Hugoniot of a single-crystal MgO preheated to 1850 K, reaching temperatures up to 9100 K and providing a complete EOS for pressures below 250 GPa~\cite{Fatyanov2018}. The principal Hugoniot of MgO has been explored in the solid and in he liquid phases using plate impacts up to 1200~GPa~\cite{Root2015}. More extreme conditions have been achieved with laser-driven shocks~\cite{McCoy2019}, providing measurements of the reflectivity, specific heat, and Gr{\"u}neisen parameter up to 2300 GPa and 64\,000 K. According to these experiments,  the commonly used tabulated EOSs, such as SESAME and LEOS, have underpredicted the temperature for a given pressure in liquid MgO. Although the melting curve of MgO has been explored with first-principles simulations up to 4000 GPa~\cite{Taniuchi2018}, the properties of the liquid have not been studied for temperatures higher than a few thousand K~\cite{Alfe2005,DeKoker2009}, where significant ionization can take place and an accurate description is required in order to interpret the latest high-temperature experiments on MgO. 
Recently, the properties of the MgSiO$_3$ plasma have been reported in the warm-dense matter regime~\cite{GonzalezMilitzer2019}, where it was shown that  the K-shell ionization of the oxygen atoms is the cause of a compression maximum along the principal Hugoniot of MgSiO$_3$. This provides an insight into the properties of silicates in the warm-dense matter regime, but a full picture is far from complete. Therefore, a consistent EOS that spans a wide range of temperatures and pressures is needed in order to provide a quantitative picture that accounts for the changes in MgO properties with pressure and temperature.

Path integral Monte Carlo (PIMC) methods have gained considerable interest as a state-of-the-art, stochastic first-principles technique for computing the properties of interacting quantum systems at finite temperature. This formalism results in a highly parallelizable implementation and an accurate description of the properties of materials at high temperature where the electrons are excited to a significant degree~\cite{Mi06,Benedict2014,Driver2015,Hu2016,ZhangBN2019}. 
The application of the PIMC method to first and second-row elements has been possible due to the development of free-particle~\cite{Ce95,Ce96} and Hartree-Fock nodes~\cite{MilitzerDriver2015}. The latter approach enables one to efficiently incorporate localized electronic states into the nodal structure, which extends the applicability of the path integral formalism to heavier elements and lower temperatures~\cite{ZhangSodium2017,Driver2018}. Furthermore, PIMC treats all electrons explicitly, avoiding the use of any pseudopotentials. The PIMC simulation time scales as $1/T$, proportional to the length of the paths, which is efficient at high-temperature conditions, where most electrons including the K shell are excited. The accuracy of PIMC at intermediate temperatures has been shown~\cite{Mi09,ZhangCH2018} to be in good agreement with predictions from density functional theory molecular dynamics (DFT-MD) simulations.

Kohn-Sham DFT~\cite{Hohenberg1964,Kohn1965} is a first-principles simulation method that determines the ground state of quantum systems with high efficiency and reasonable accuracy, which has made it a workhorse of computational materials science. It has thus
gained considerable use for many years. The introduction of the Mermin
scheme~\cite{Mermin1965} enabled the inclusion of excited electronic
states, which extended the applicability range of the DFT
method to higher temperatures. The combination of this method with
molecular dynamics has been widely applied to compute the EOS of
condensed matter, warm dense matter (WDM), and dense
plasmas~\cite{Root2010,Wang2010,Mattsson2014,Zhang2018}. It is often
the most suitable computation method to derive the EOS under a wide
range of conditions because it accounts for both electronic shell and
bonding effects. The main source of uncertainty in DFT is the use of an approximate exchange-correlation (XC) functional. The error in the XC functional is generally a negligible fraction of the internal energy, which is the most relevant quantity for the EOS and the derivation of the shock Hugoniot curve ~\cite{Karasiev2016}. However,
the range of validity of this assumption in the WDM regime remains to 
be verified for different classes of materials through the comparison 
with laboratory experiments and other simulation methods like PIMC.

In this work, we combine the PIMC and DFT-MD methods to study the
properties of MgO in the regime of WDM. The combination of both
methods allows us to test the validity of the approximations in the
methods. We study the regimes of thermal and pressure ionization of
the electronic shells and provide an equation of state that spans a
wide range of temperatures and pressure.  We describe the electronic
properties of liquid MgO and show how the band gap between s and p
states closes upon compression, and provide a structural
characterization of the liquid.  Finally, we compare our PIMC/DFT-MD
shock Hugoniot curves with widely used models and experiments.

\section{Simulations Methods}

\begin{figure}
    \centering
    \includegraphics[width=\columnwidth]{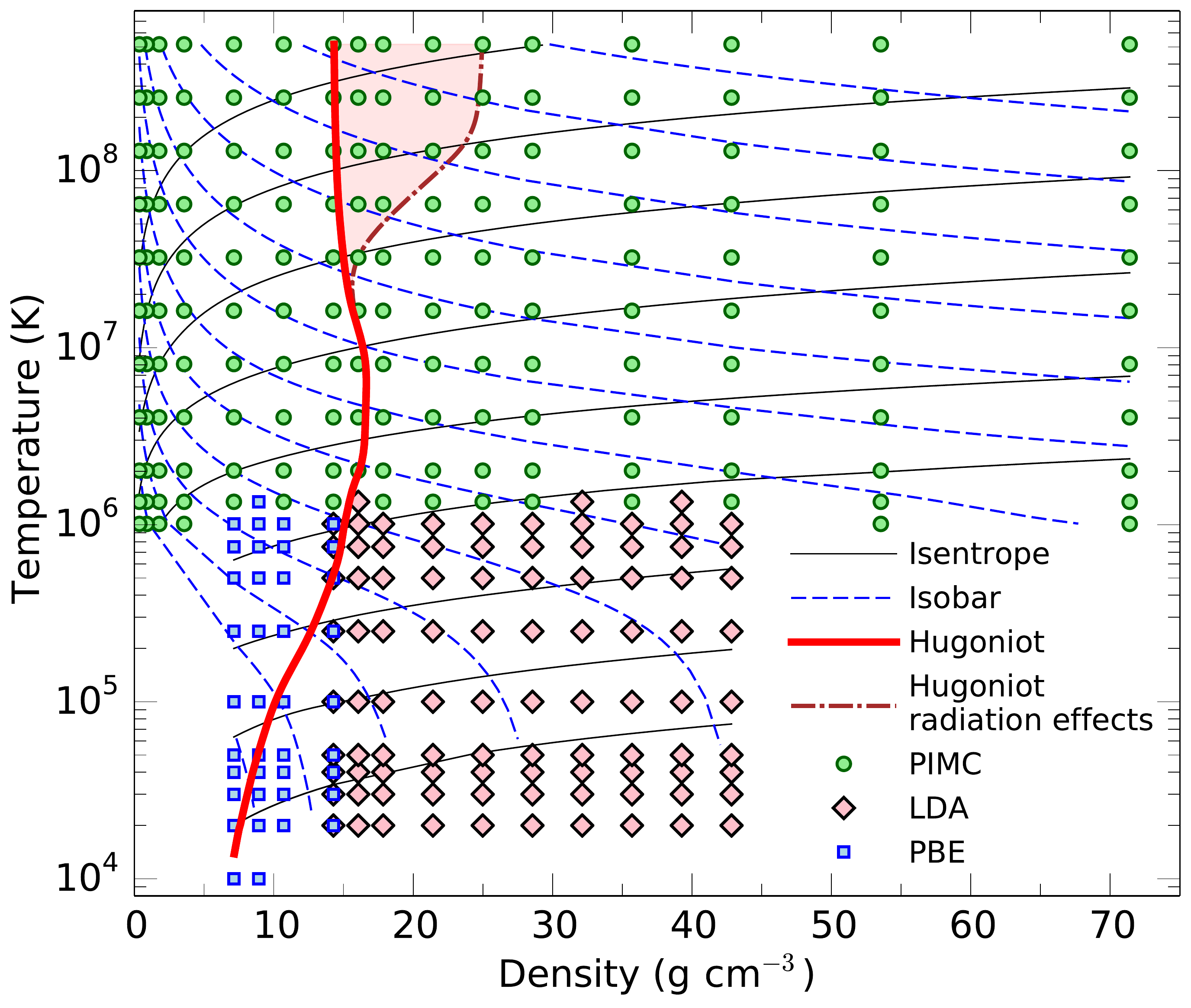}
    \caption{Temperature-density conditions of our PIMC and DFT-MD simulations along with
computed isobars, isentropes and shock Hugoniot curves that were with and without radiation effects for an initial density of
$\rho_0=3.570$ g cm$^{-3}$ ($V_0=18.7478216$~\AA/f.u.).
\label{fig:Tvsrho}}
   \label{fig:Hugoniot}
\end{figure}

We perform first-principles computer simulations of magnesium oxide for a range of extreme density and temperature conditions that we illustrate in Fig.~\ref{fig:Tvsrho}. At high temperature, we employ path integral Monte
Carlo (PIMC) simulations and at lower temperatures we use standard Kohn-Sham density functional theory molecular
dynamics (DFT-MD) calculations. 

\subsection{PIMC}

PIMC is a state-of-the-art
first-principles technique for simulating interacting
quantum systems at finite temperature. The fundamental techniques for the simulations of bosonic systems
were developed in Ref.~\cite{PC84} and later
reviewed in Ref.~\cite{Ce95}. Subsequently the algorithm was
extended to simulate fermionic systems by introducing the {\em restricted} paths approach~\cite{Ce91,Ce92,Ce96}. The first results of this simulation method were
reported in the seminal work on liquid $^3$He~\cite{Ce92} and dense
hydrogen~\cite{PC94}. In subsequent articles, this method was applied to
study hydrogen~\cite{Ma96,Mi99,MilitzerThesis,MC00,MC01,Mi01},
helium~\cite{Mi06,Mi09,Mi09b}, hydrogen-helium mixtures~\cite{Mi05}
and one-component plasmas~\cite{JC96,MP04,MP05}. In recent years, the
PIMC method was extended to simulate plasmas of various first-row
elements~\cite{Benedict2014,DriverNitrogen2016,Driver2017,ZhangCH2017,ZhangCH2018,Zhang2018}
and with the development of Hartree-Fock nodes, the simulations of
second-row elements became
possible~\cite{MilitzerDriver2015,Hu2016,ZhangSodium2017,Driver2018}.

This PIMC method is based on the thermal density matrix of a quantum
system, $\hat\rho=e^{-\beta \hat{\cal H}}$, that is expressed as a
product of higher-temperature matrices by means of the identity
$e^{-\beta \hat{\cal H}}=(e^{-\tau \hat{\cal H}})^M$, where
$M$ is an integer and $\tau\equiv\beta/M$ represents the time step of a path integral in
imaginary time. The path integral emerges when the operator $\hat\rho$
is evaluated in real space,
\begin{equation}
\left<\mathbf R|\hat\rho| \mathbf R'\right>=\frac{1}{N!}\sum_{\mathcal P}(-1)^{\mathcal P}\oint_{\mathbf R\to\mathcal P\mathbf R'}\mathbf{dR}_t\, e^{-S[\mathbf R_t]}.
\label{PI}
\end{equation}
Here, we have already summed over all permutations, $\mathcal P$, of
all $N$ identical fermions in order project out all antisymmetric
states.  For sufficiently small time steps, $\tau$, all many-body
correlation effects vanish and the action, $S[\mathbf R_t]$, can be
computed by solving a series of two-particle
problems~\cite{PC84,Na95,BM2016}. The advantage of this approach is that all many-body quantum correlations are recovered
through the integration over all paths. The integration also enables
one to compute quantum mechanical expectation values of thermodynamic
observables, such as the kinetic and potential energies, pressure,
pair correlation functions and the momentum
distribution~\cite{Ce95,Militzer2019}. Most practical implementations of
the path integral techniques rely on Monte Carlo sampling techniques
because the integral has $D \times N \times M$ dimensions ($D$ is the dimension in space) and, in
addition, one needs to sum over all permutations. The method becomes
increasingly efficient at high temperature because the length
of the paths scales like $1/T$. In the limit of low temperature, where
few electronic excitations are present, the PIMC method becomes
computationally demanding and the MC sampling can become inefficient.
Still, the PIMC method avoids any exchange-correlation approximation
and the calculation of single-particle eigenstates, which are embedded 
in all DFT calculations. 

The only uncontrolled approximation within fermionic PIMC calculations
is the use of the fixed-node approximation, which restricts the paths
in order to avoid the well-known fermion sign
problem~\cite{Ce91,Ce92,Ce96}. Addressing this problem in PIMC is
crucial, as it causes large fluctuations in computed averages due to
the cancellation of positive and negative permutations in
Eq.~\eqref{PI}. We solve the sign problem approximately by restricting
the paths to stay within nodes of a trial density matrix that we obtain 
from a Slater determinant of single-particle density matrices,
\begin{equation}
\rho_T({\bf R},{\bf R'};\beta)=\left|\left| \rho^{[1]}(r_{i},r'_{j};\beta) \right|\right|_{ij}\;,
\label{FP}
\end{equation}
that combined free and bound electronic states~\cite{MilitzerDriver2015,Driver2018}, 
\begin{eqnarray}
\label{rho1}
\rho^{[1]}(r,r';\beta) &=& \sum_{k} e^{-\beta E_k} \, \Psi_k(r) \, \Psi_k^*(r')\\
 &+& \sum_{I=1}^{N} \sum_{s=0}^{n} e^{-\beta E_s} \Psi_s(r-R_I) \Psi_s^*(r'-R_I)\;.
\quad.
\end{eqnarray}
The first sum includes all plane waves, $\Psi_k$ while the second represents $n$ 
bound states $\Psi_s$ with energy $E_s$ that are localized around all atoms $I$. Predictions from various slightly differing forms of this approach have been compared in Ref.~\cite{ZhangSodium2017} 

All PIMC simulations used periodic boundary conditions and treated 80 electrons, 4 Mg and 4 O nuclei explicitly. We enforced the nodal constraint in small steps of imaginary time of $\tau=1/8192$ Ha$^{-1}$, while
the pair density matrices~\cite{Militzer2016} were evaluated in steps of 1/1024 Ha$^{-1}$. This
results in using between 2560 and 5 time slices for the temperature
range that is studied with PIMC simulations here. These choices converged
the internal energy per atom to better than 1\%.  We have shown the
associated error is small for relevant systems at sufficiently high
temperatures~\cite{Driver2012}.

\subsection{DFT-MD}

\begin{figure}[!t]
    \centering
    \includegraphics[width=\columnwidth]{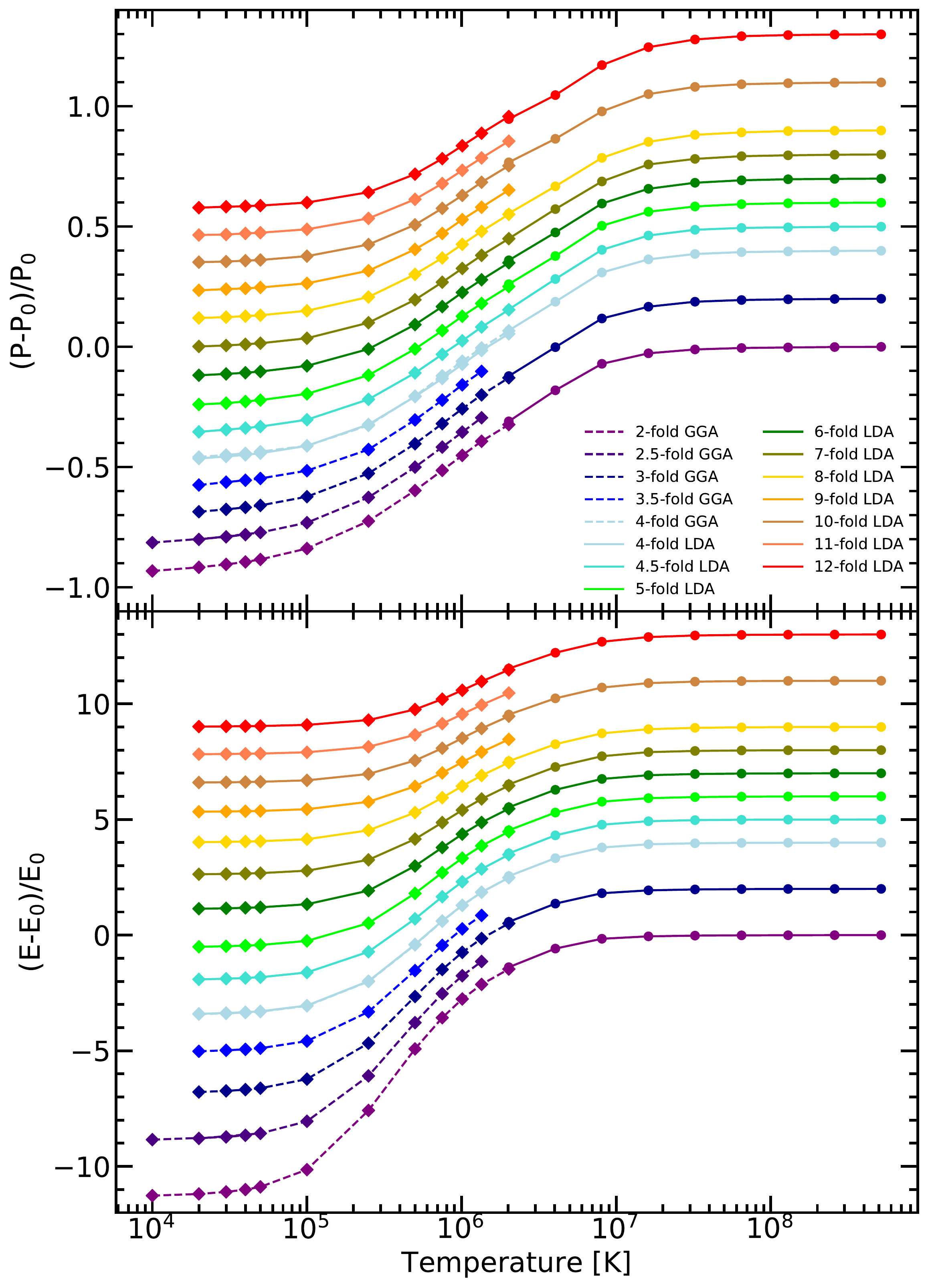}
    \caption{Reduced equation of state of MgO from 2 to 12-fold compression from ambient density as a function of temperature as predicted by DFT (diamonds) and PIMC (circles). The top panel is the relative difference between the total pressure $P$ and the pressure $P_0$ of an ideal gas of nuclei and a Fermi gas of electrons. The bottom panel is the difference between the total internal energy $E$ and the energy $E_0$ of an ideal gas of nuclei and a Fermi gas of electrons. Different isochores have been shifted apart for clarity.}
    \label{fig:EOS}
\end{figure}

Kohn-Sham (KS) DFT-MD~\cite{Hohenberg1964,Kohn1965} is a well-stablished method to compute the properties of matter in the warm dense matter regime. We thus used the DFT-MD code VASP \cite{Kresse1999} to perform simulations up to 2 millions Kelvin to complement the PIMC calculations. We restricted our DFT-MD calculations to a range of 2- to 12-fold the ambient density of MgO. We placed from 4 to 32 MgO formula units in a cubic box with periodic boundary conditions. It has been shown in previous work that such a small cell is not detrimental to the accuracy of the EOS data under such high temperatures \cite{Driver2015b,ZhangCH2017,Driver2018}. To keep the temperature constant, we used a Nos\'e thermostat  \cite{Nose1984,Nose1991}. The timestep was adapted to the density and the temperature and ranged from  8.33~as to 0.25~fs for a total duration of several hundred time steps to ensure a relevant estimation of the thermodynamic quantities.

Regarding the DFT calculations, we used the Mermin scheme \cite{Mermin1965} and employed projector augmented wave \cite{Blochl1994} pseudopotentials. We were however limited to the existing pseudopotentials of VASP and opted for hard ones with a 1s$^2$ frozen core for both magnesium and oxygen. The PAW sphere radius was 1.75 bohr for Mg and 1.1 bohr for O. We used the generalized gradient approximation Perdew-Burke-Ernzerhof (PBE) \cite{PBE} functional for the lowest densities as it has shown to give good results for MgO \cite{Soubiran2018}. At high density the pseudopotentials in PBE were unable to give proper results and we switched to the local density approximation (LDA). We have a very good agreement between both functionals at 4-fold compression (see paragraph \ref{sec:EOS}). For very high temperatures, the Mermin approach requires computation of the excited states even for low occupation numbers. That is why we computed up to 6000 bands even for the reduced cell size of 8 atoms when we reached temperatures above 10$^6$~K. This high number of bands ensured that we computed all the bands up to an energy level associated to an occupation of 10$^{-5}$. The systematic error due to this band cut-off is below 0.1\%. The energy cut-off for the plane wave basis set had also to be increased at high temperatures up to 7000~eV. We sampled the Brillouin zone with the $\Gamma$ point only which was found to be sufficient for the convergence of the thermodynamic quantities under the conditions of interest.

For the calculation of the density of state (DOS), we used the KS eigenenergy on snapshots sampled every 500 time steps to compute an instantaneous DOS. To obtain a smooth curve, we replaced each eigenvalue by a gaussian centered on the eigenenergy and with an RMS width of 0.5~eV. Each instantaneous DOS was aligned on its Fermi energy so that we could average them. The average DOS was the re-shifted by the average value of the Fermi energy. We also computed the projected DOS (pDOS). Each eigenstate was locally projected on spherical harmonics centered on each nucleus. We could then average the contributions to the DOS by species and by harmonics level.

\section{Results and discussion}

\subsection{Equation of state \label{sec:EOS}}
\begin{figure}[!th]
    \centering
    \includegraphics[width=\columnwidth]{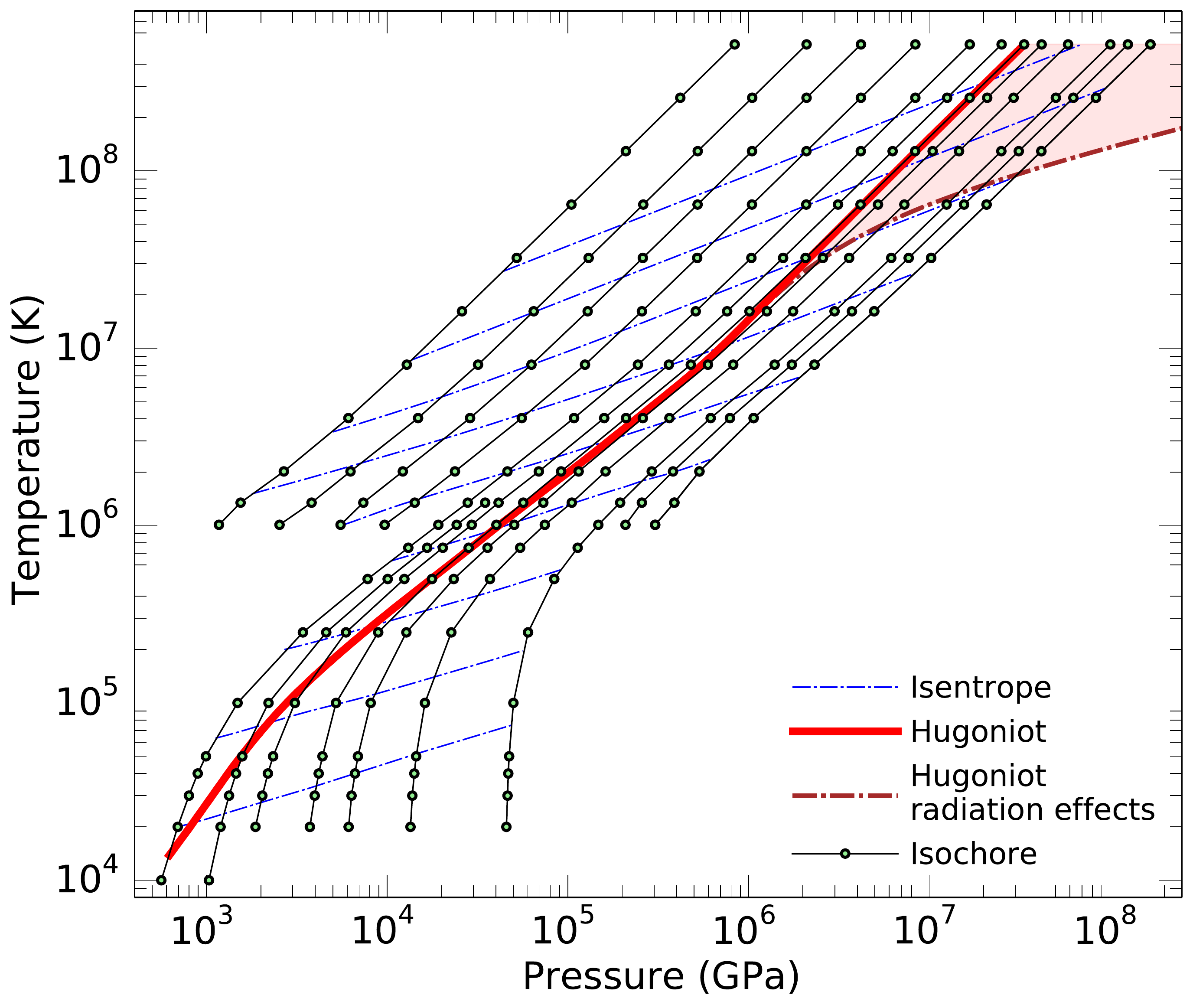}
    \caption{Temperature-pressure conditions for the PIMC and DFT-MD
    calculations along isochores corresponding to the densities of
    0.1-fold (upper left curve) to 20-fold (right curve) the ambient density. The shock Hugoniot curves with and without radiation effects were included as well as a number of isentropes.
  }\label{fig:PvsT} 
\end{figure}

The internal energies that were derived in the VASP DFT-MD simulations with LDA and PBE functionals were shifted respectively by $-273.6020398$ and $-274.9006146$~Ha per MgO formula unit in order to make them compatible with the all-electron energies that we obtained with the PIMC simulations. 

In Fig.~\ref{fig:EOS} we plotted the relative excess pressure and energy with respect to an ideal gas of nuclei and Fermi gas of electrons. Regarding the DFT part solely, there is good agreement between LDA and PBE calculations at 4-fold compression ensuring a smooth transition from one data set to another. As temperature increases, DFT becomes too computationally expensive around the same conditions that PIMC first becomes viable. For MgO we managed to obtained good agreement between PIMC and DFT at 1,347,305~K. Near this temperature, the relative difference in the pressure is less than 3.5\%, and the difference in the energy ranges from 2.5 to 5.3~Ha/nucleus. These are maximum differences between PIMC and DFT-MD, as they are reported for the 12-fold compression case, where we are most severely affected by pseudo-potential overlap and frozen core effects. Globally, the agreement is very satisfactory. 

With our PIMC simulations, we were able to reach densities as low as 0.1 times the ambient density but not with DFT because of numerical limitations (both in computer time and in memory needs). For high densities, DFT was limited by the pseudopotentials available, preventing any simulations beyond 12-fold compression. However, as can be seen in Fig.~\ref{fig:PvsT} our data set still covers an extremely wide range of pressures and temperatures. 
Our entire EOS table is provided as supplementary material. 
This allows to benchmark other faster but more approximate EOS methods for a wide range of conditions. 

To extrapolate the present EOS towards other conditions it is possible to use other approximations such as the fully ionized plasma \cite{Chabrier1998} for high densities or a chemical model at low density. Such an approach has been recently performed on H-He mixtures \cite{CMS2019}. It is also possible to improve the EOS by a careful characterization of the free energy using thermodynamic integration. This work has been performed on H-He at temperatures relevant for the interior of giant planets \cite{Militzer2013b}.

\subsection{Electronic and ionic structures}
\begin{figure*}[!hbt]
    \centering
    \includegraphics[width=17.5cm]{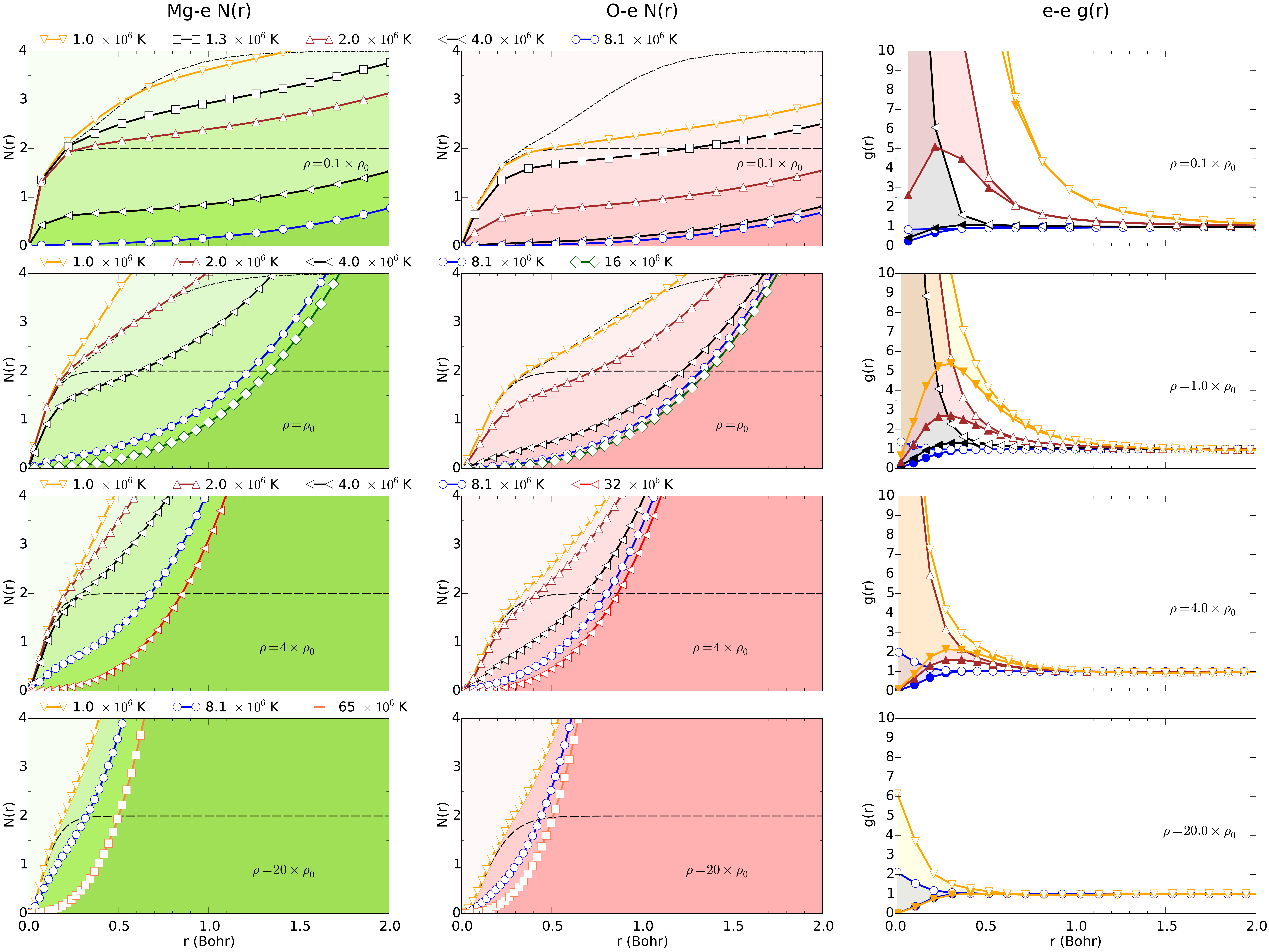}
    \caption{  The left and middle columns show the integrated nuclear-electron pair correlation, functions $N(r)$, for Mg and O nuclei respectively. In the right column, we plot the electron-electron correlation functions, $g(r)$, for pair of electrons with parallel ({filled} symbols) and anti-parallel spins ({open} symbols). All curves were derived from PIMC simulations. The four rows show results for different densities of ranging from 0.1, 1.0, 4.0, and 20.0 times the ambient density of 3.569856 g cm$^{-3}$. The $N(r)$ functions represent the average of number of
  electrons contained within a sphere of radius, $r$, around a given nucleus. For comparison we show the corresponding functions with thin dashed lines for isolated nuclei with double occupied 1s core states
  that we computed with the GAMESS software~\cite{GAMESS}. For the lowest two densities, the thin dash-dotted lines show the curve for double occupied 1s and 2s states.  
  \label{fig:N(r)}
  }
\end{figure*}

\begin{figure}[!t]
    \centering
    \includegraphics[width=0.91\columnwidth]{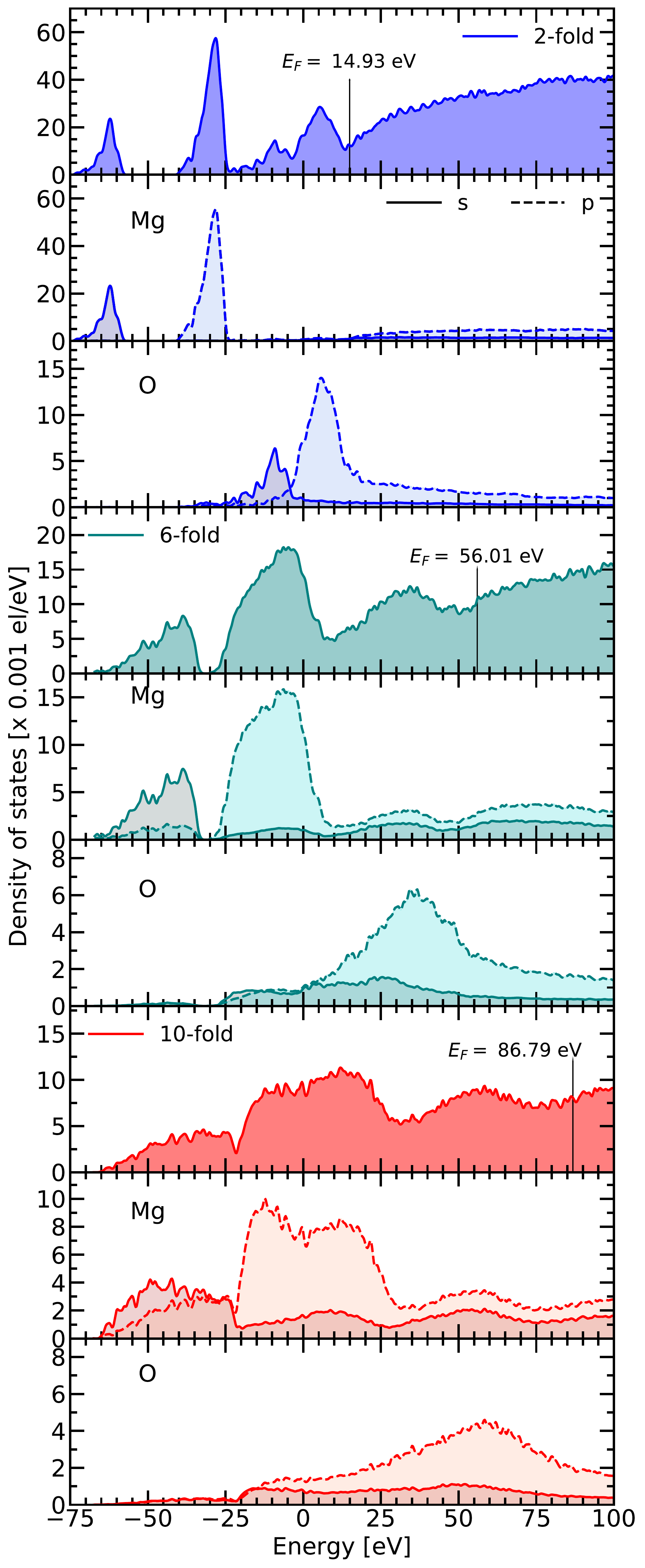}
    \caption{Electronic density of states of MgO at 100~000~K. They are plotted for 2-fold (first three panels in blue), 6-fold (middle panels in green) and 10-fold (last three panels in red) compression. For each compression, we plotted the full DOS in the first panel, the projected DOS on s (solid) and p (dashed) orbitals of magnesium in the second panel and the projected DOS on s and p orbitals of oxygen in the third panel. The DOS are aligned to the zero of energy of VASP. The vertical lines indicate the Fermi level.}
    \label{fig:DOS}
\end{figure}

From PIMC simulations, a measure of the degree of ionization can be
obtained from the integrated nucleus-electron pair correlation
function, $N(r)$, given by
\begin{equation}\label{eq:N(r)}
N(r)= \left<\frac{1}{N_I}\sum_{e,I}\Theta(r -\|\vec r_e-\vec r_I \|)\right>,
\end{equation}
where $N(r)$ represents the average number of electrons within a sphere of
radius $r$ around a given nucleus of atom of type $I$.  The summation includes
all electron-nucleus pairs and $\Theta$ represents the Heaviside function.
Fig.~\ref{fig:N(r)} shows the integrated nucleus-electron pair correlation
function for temperatures from $1\times 10^6$ K to $65\times 10^6$ K and
densities from 0.357  g$\,$cm$^{-3}$ (0.1-fold) to 71.40 g$\,$cm$^{-3}$ (20-fold
compression). For comparison, the $N(r)$ functions of an isolated nucleus with a
doubly occupied 1s orbital were included. Unless the 1s state
is ionized, its contribution will dominate the $N(r)$ function at small
radii of $ r < 0.2 $ Bohr radii. For larger radii, contributions from other
electronic shells and neighboring nuclei will enter. Still, this is the most
direct approach available to compare the degree of 1s ionization of the three
nuclei. 

At 0.1-fold compression, the comparison with the corresponding curves for the
isolated nuclei shows that the ionization of the 1s states of the Mg
nuclei occurs over the temperature interval from 2.0 to $8.1\times 10^6$ K.
Conversely, the ionization of 1s state of the oxygen nuclei occurs from 
1.3 to 4.0 $\times 10^6$ K, which reflects the difference in binding energy that
scales with the square of the nuclear charge, $Z$. Consistent with this
interpretation, one finds no evidence of an ionization of the 2s states of Mg 
nuclei at 1.0 $\times 10^6$ K while those of the oxygen nuclei are partially 
ionized, as the comparison with calculations of isolated nuclei in 
Fig.~\ref{fig:N(r)} shows.

When the density is increased from 0.1- to 1.0-fold compression
(second row of panels in Fig.~\ref{fig:N(r)}), the degree of 1s ionization is
reduced. For the Mg and O nuclei, the $N(r)$ functions at small $r$ are
closer to doubly occupied 1s state than they were before. This trend
continues as we increase the density to 4.0 and 20-fold
compression. The degree of 1s ionization is consistently reduced
with increasing density
when the results are compared for the same temperature. Most noticeable 
are the deviations for a temperature of $8.1 \times 10^6$ K. At 0.1-fold
compression the 1s states of both nuclei are essentially fully ionized 
while they still appear to be partially occupied for this temperature at 20-fold compression.

In Fig.~\ref{fig:N(r)}, we also show the electron-electron pair
correlation functions, $g(r)$, that we derived from our all-electron
PIMC simulations. Without any Coulomb interactions, this function
would equal 1 for all $r$ for electron pairs with opposite spin. This
is also true for pairs with the same spin for large distances, but for
small $r$, this function must drop to zero because of Pauli
exclusion. The trend for very small $r$ can be seen in right column of
Fig.~\ref{fig:N(r)} where we plot the $g(r)$ functions for the fully
interacting MgO system. For all other $r$ values, these functions are
substantially modified by the Coulomb interaction effects, as
expected. At low density and low temperature, one finds strong
positive correlations that reflect the fact that electrons with both
types of spins occupy bound states at the same nuclei. With increasing
temperature, the electrons become ionized and the positive correlation
disappears. At small $r$ Pauli exclusion dominates which introduces
difference into the pair correlation functions between electrons with
parallel and opposite spin. For large $r$, these function are always
very similar.

For lower temperature conditions that we studied with DFT-MD simulations, one can 
derive information about the electronic properties by analyzing the Kohn-Sham eigenstates. In
Fig.~\ref{fig:DOS} we plotted the density of states (DOS) for 10$^5$~K
at three different densities. The full DOS shows that the
system becomes increasingly pressure ionized with increasing density. 
Peaks in the DOS, that are narrow at low density,
broaden and merge with the continuum of free states. In the KS-DFT
framework, we can also approximately decompose the DOS into contributions 
from different nuclei and orbital character. We included the results 
from various such projections in Fig.~\ref{fig:DOS}.
At 2-fold compression, the first two peaks
correspond to the L shell of magnesium while the M-shell states have already merged 
with the continuum. The second set of two
peaks between $-$20 and 20~eV correspond to the s and p orbitals of oxygen L-shell. We
retrieve a very similar feature in our simulations for pure oxygen~\cite{Driver2015b} 
that showed oxygen orbitals that could be distinguished clearly from the continuum.

As the density is increased, we can still identify two L-shell peaks of
magnesium: a first one (L$_\textrm{I}$) with s character and a second 
(L$_\textrm{II,III}$) with p character. However, already at 6-fold and much more at
10-fold compression, both types of orbitals hybridize and form very broad peaks
without a clear gap. For oxygen, the s-type peak has already
disappeared at 6-fold which is very different from predictions for pure oxygen~\cite{Driver2017b}. However, 
it reminds one of the electronic properties of the carbon in dense plastics~\cite{ZhangCH2017}. 
This phenomena is likely caused by the hybridization
of the p orbital of magnesium with the s and p orbitals of oxygen. This produces
a very large broadening of the oxygen L-band which gradually merges with the
continuum with increasing density. The observed hybridization of magnesium and oxygen orbitals has
thus consequences for the ionization state of MgO at extreme conditions. These effects will be difficult to 
reproduce with average atom models that describe
mixtures of dense atoms as a collection of noninteracting species.
 

\begin{figure*}[!hbt]
    \centering
    \includegraphics[width=17.5cm]{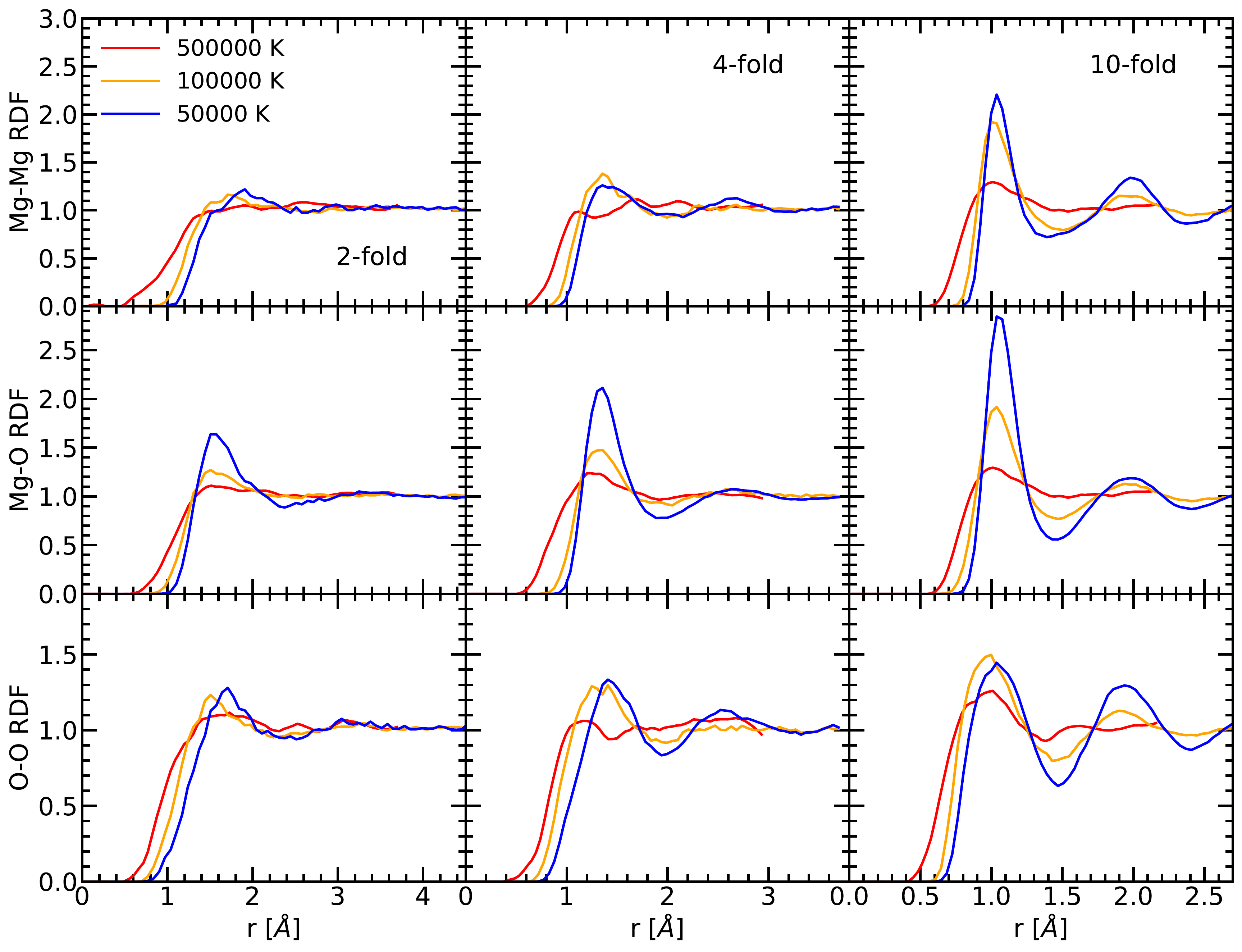}
    \caption{ Nuclear pair correlation functions 2-fold (left), 4-fold (middle) and 10-fold compression (right column) at different temperatures given in the legend. The different rows show these functions for Mg-Mg, Mg-O, and O-O pairs of nuclei.
  \label{fig:g(r)}
  }
\end{figure*}

With the nuclear configurations extracted from the MD trajectories, we
can determine the local structure of the plasma as a function of the
density and temperature. In Fig.~\ref{fig:g(r)}, we illustrate changes 
in this structures with a series of pair correlation functions. With increasing 
density and decreasing temperature, the liquid becomes more structured and the 
peaks in the correlation functions increase. Under no conditions we found a 
signature of a stable molecular bond between Mg and O species. 
However, the Mg and O nuclei exhibit positive correlations that are slightly stronger than 
that between the other pairs. This was seen in Ref.~\cite{Cebulla2014} and is due to
an imbalance in their respective charge states. 
In comparison to our recent work~\cite{ZhangBN2019} on boron nitride (BN) which
shows a structure-less feature in $g(r)$ at $5\times10^4$--$10^5$ K,
our results on MgO show that correlation remains up to 5$\times10^5$ K. 
This is consistent with the electronic structure analysis above and indicates that an average-atom approach would be
less trustworthy for MgO at below 5$\times10^5$ K,
higher than that for BN ($10^5$ K) because of the deeper K shells of Mg and O than B and N.

\subsection{Single and Multi Shock Hugoniot Curves}

Dynamic shock compression experiments allow one to directly measure
the equation of state and other physical properties of hot, dense
fluids.  Such experiments are often used to determine the principal
Hugoniot curve, which is the locus of final states that can be
obtained from different shock velocities. Density functional theory
has been validated by experiments as an accurate tool for predicting
the shock compression of a variety of different
materials~\cite{Root2010,Wang2010,Mattsson2014}.

During a shock wave experiment, a material whose initial
state is characterized by an internal energy, pressure, and volume,
($E_0,P_0,V_0$), will change to a final state denoted by $(E,P,V)$
while conserving mass, momentum, and energy. This leads to the
Rankine-Hugoniot relation~\cite{Ze66},

\begin{equation}
(E-E_0) + \frac{1}{2} (P+P_0)(V-V_0) = 0.
\label{hug}
\end{equation}

\begin{figure}
    \centering
    \includegraphics[width=\columnwidth]{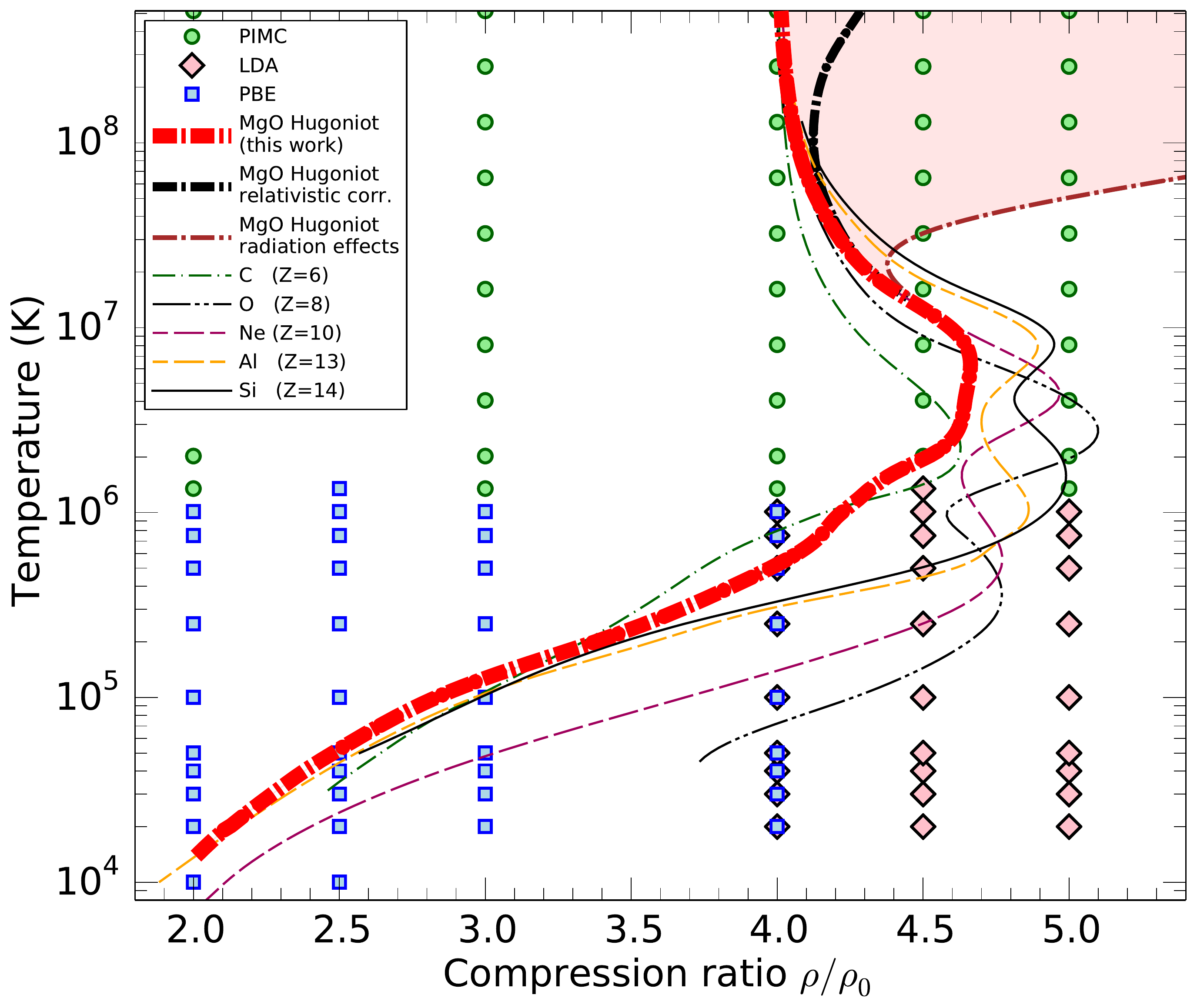}
    \caption{The MgO shock Hugoniot curve with and without relativistic and radiation effects are compared with the Hugoniot curves of carbon~\cite{Driver2012,Benedict2014},
oxygen~\cite{Driver2015b},
neon~\cite{Driver2015}, aluminum~\cite{Driver2018}, and
silicon~\cite{MilitzerDriver2015,Hu2016}. This comparison supports the interpretation that the broad temperature interval where the compression of MgO exceeds 4.5 can be attributed to the ionization of the K shell electrons of the oxygen and magnesium nuclei.}
    \label{fig:HugoniotComp}
\end{figure}

\begin{figure}
    \centering
    \includegraphics[width=\columnwidth]{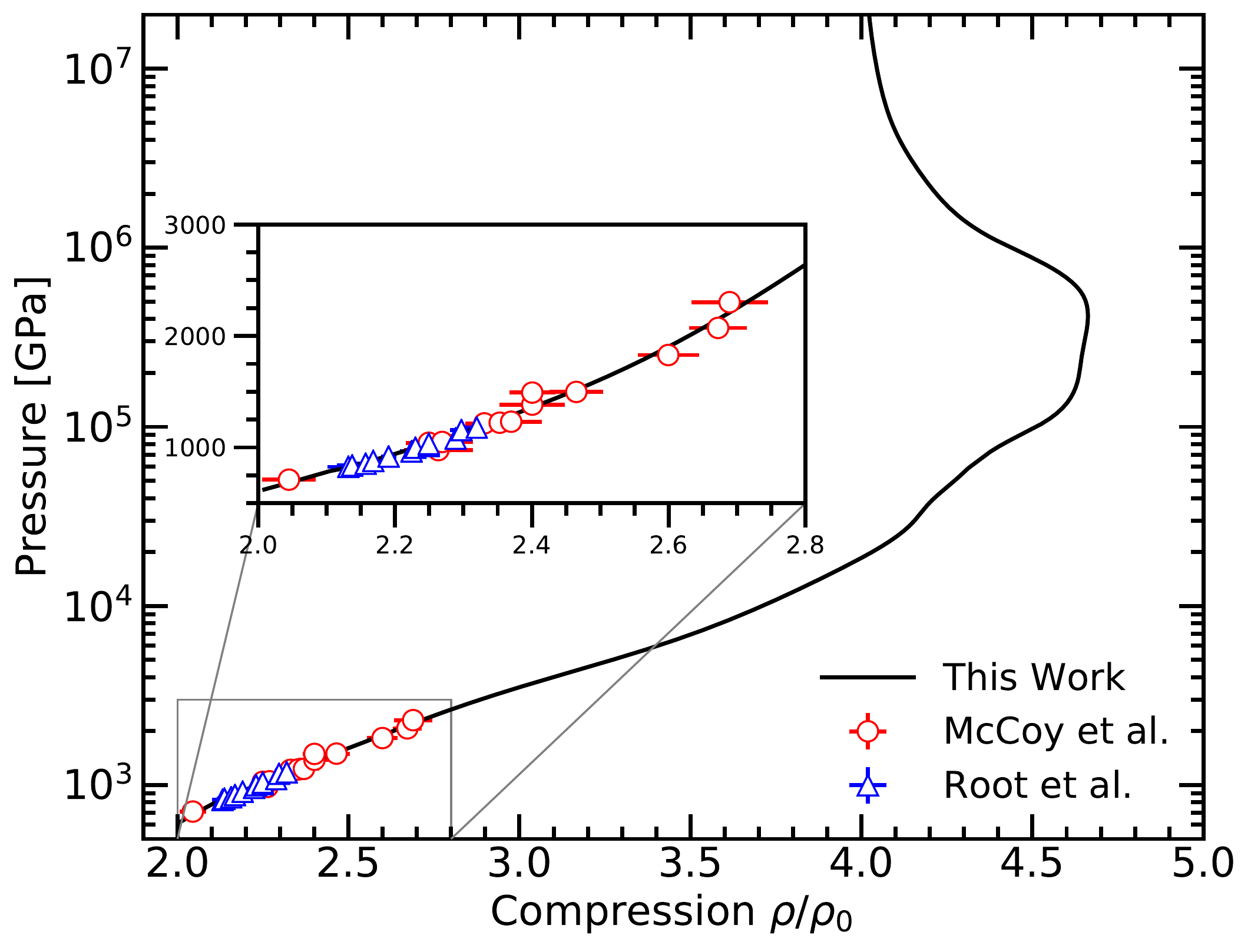}
    \caption{The MgO shock Hugoniot curve in the pressure-compression plane. The symbols show the experimental results by Root \textit{et al.} \cite{Root2015} and by McCoy \textit{et al.} \cite{McCoy2019}. The inset enlarges the lower pressure region.}
    \label{fig:HugoniotP}
\end{figure}

\begin{figure}
    \centering
    \includegraphics[width=\columnwidth]{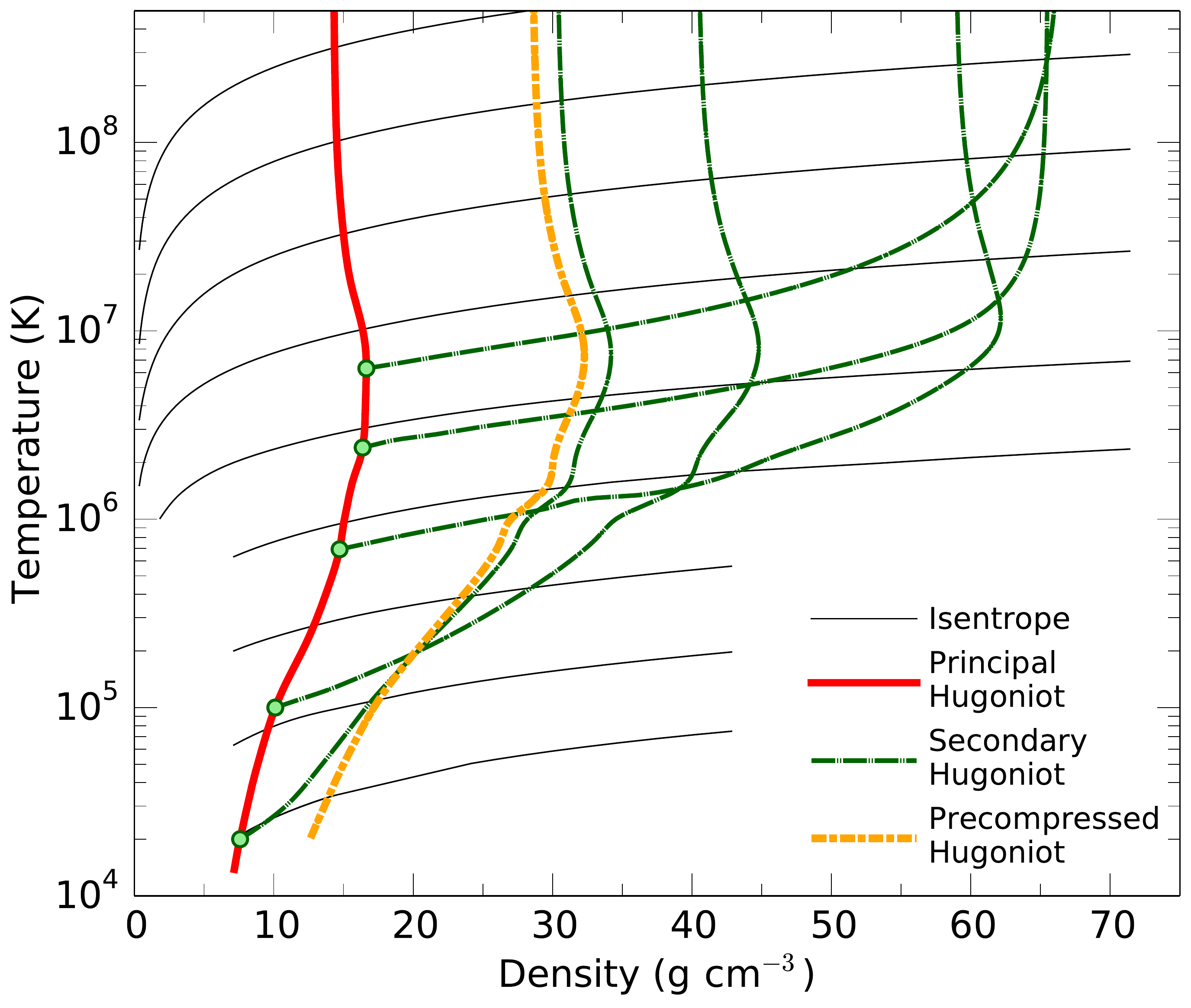}
    \caption{Single and double shock Hugoniot curves are compared with a family of isentropes. For weak second shocks, minimal shock heating is introduced and the secondary Hugoniot curves behave like isentropes. For stronger second shocks, the heating is substantial and the resulting Hugoniot curve is similar to the shock curve for a precompressed sample.}
    \label{fig:DoubleShock}
\end{figure}

Here, we solve this equation for our computed first-principles EOS
data set, only in the liquid phase. We obtain a continuous Hugoniot curve by
interpolating the EOS data with splines as a
function of $\rho$ and $T$. The Hugoniot curve for the liquid state only depends on the liquid EOS and the initial point, but not on the path followed in the solid phases. In order to obtain the principal Hugoniot curve, we used initial conditions based on the energy and pressure of ambient,
solid MgO in the B1 phase computed with static DFT calculations at the experimental density of 3.570 g$\,$cm$^{-3}$ ($V_0=$18.7478216 $\rm \AA^3$/MgO). Depending on which DFT functional we use, we obtained two slightly different initial energies, $E_0^{\rm LDA} = -274.6138119$ and $E_0^{\rm PBE} = -275.3370951$ Ha/MgO. The difference between these two values is small compared the $\sim$50$\,$000 Ha/MgO that the internal energy changes along the shock Hugoniot curve in the temperature interval that we study here. However, one would want to make reasonable choice to minimize the error that arises from choosing a particular DFT functional. Thus, when we evaluate the $(E-E_0)$ term in Eq.~\ref{hug}, we use values derived with the same functional. When we use PIMC values for $E$, we combine them with $E_0^{\rm PBE}$ because this has work well in Ref.~\cite{ZhangSodium2017}. The resulting shock Hugoniot curve has been added to Figs.~\ref{fig:Hugoniot}, \ref{fig:PvsT}, \ref{fig:HugoniotComp}, \ref{fig:HugoniotP}, and \ref{fig:DoubleShock}.

The principal Hugoniot curve in Fig.~\ref{fig:HugoniotComp} and \ref{fig:HugoniotP} exhibits a wide
pressure interval where the compression ratio exceeds 4.0, the asymptotic value
for a non-relativistic ideal gas. Such high compression values are the result of
excitations of internal degrees of freedom~\cite{Mi06}, which
increase the internal energy term in Eq.~\eqref{hug}. Consequently, the
second term in this equation becomes more negative, which reduces the
volume $V$ and thus increases the compression ratio. Indeed, at a pressure of
18666 GPa and a temperature of 525$\,$000$\,$K, the shock compression
ratio starts to exceed 4, which are conditions where the $L$ shell
electrons are already ionized. The bulk of the high compression region is
dominated by the ionization of the K shell (1s) electrons of the Mg and O
nuclei. However, instead of two separate compression peaks, we see a broad 
region of increased compression indicating that the two peaks have merged 
due to interaction effects~\cite{Mi09}. No experimental data are available in the K- and L-shell ionization regimes yet. In the TPa regime, we obtain an excellent agreement between our predictions and the experimental data by Root {\it et al.}~\cite{Root2015} and McCoy {\it et al.}~\cite{McCoy2019} (see Fig.~\ref{fig:HugoniotP}), which underline the robustness of our DFT-MD calculations.

The highest compression ratio of 4.66 is reached for 6.73$\times 10^7$
K and 421$\,$000 GPa, which coincides in pressure with the upper
compression maximum of the shock Hugoniot curve of pure silicon, which
has also been attributed to K shell
ionization~\cite{MilitzerDriver2015}. Based on this comparison and the
K shell ionization analysis of MgO in Fig.~\ref{fig:N(r)} we
conclude that the upper part of the high compression region in
Fig.~\ref{fig:HugoniotComp} is dominated by the ionization of the K shell
electrons of the Mg nuclei while the lower end around
10$^5$ GPa and 2$\times$10$^6$ K, marks the beginning of
the K shell ionization of the oxygen ions as Fig.~\ref{fig:N(r)}
confirms. However, in shock compressed pure oxygen, the K shell
ionization peak occurs for lower P and T. We attribute this difference
to interaction effects in hot, dense MgO that can shift the
compression peaks along the Hugoniot curve to higher temperatures and
pressures and reduce the peak compression~\cite{Mi09}. 

It is interesting to note that we do not see a separate compression maximum for the L shell ionization while such maxima were seen in simulations of pure nitrogen and oxygen~\cite{DriverNitrogen2016,Driver2018}. In Figs.~\ref{fig:HugoniotComp} and ~\ref{fig:HugoniotP}, we instead see a slight lowering in temperature, or pressure respectively, from approximately 3.2 to 4.2-fold compression, which we attribute to the L-shell ionization of the Mg and O nuclei because the same shape was seen in simulations of dense CH plastic and associated with the ionization of the L-shell of the carbon nuclei (see Fig. 3 of Ref.~\cite{ZhangCH2017}). We conclude that L shells of the Mg and O nuclei are ionized gradually, as
it occurs in dense carbon and boron
materials~\cite{ZhangCH2018,Zhang2018,ZhangBN2019}. This interpretation is consistent with the electronic structure observed in Fig. \ref{fig:DOS} where the oxygen and magnesium electronic states are strongly hybridized at high compression. 

The Hugoniot curves in Figs.~\ref{fig:HugoniotComp} and ~\ref{fig:HugoniotP} show a small shoulder at 4.3-fold compression. We determined this to be a robust feature even though we switch between DFT-MD and PIMC data sets in this regime. Our EOS table (see supplemental material) is sufficiently dense so that different interpolation schemes for the pressure and internal energy yield consistent results. For example, even with linear interpolation we have obtained very similar Hugoniot curves.  

Very approximately, we added relativistic and radiation effects to the
Hugoniot curves in Fig.~\ref{fig:HugoniotComp}. Under the assumption of
complete ionization, the relativistic corrections were derived for an
ideal gas of electrons. This increases the shock compression ratio for
$P > 7 \times 10^6$ GPa and $T > 10^8$ K.

Considering an ideal black
body scenario, we derived the photon contribution to the EOS using
$P_\text{radiation}=(4\sigma/3c)T^4$ and
$E_\text{radiation}=3 P_\text{radiation}V$, where $\sigma$ is the
Stefan-Boltzman constant and $c$ is the speed of light in vacuum.
We find that radiation effects are important only for temperatures
above $2\times10^7$ K, which is above the temperature necessary to
completely ionize the 1s orbitals of Mg and O species.


In Fig.~\ref{fig:DoubleShock}, we show a number of double-shock Hugoniot
curves. Various points along the principal Hugoniot curve were chosen
as initial conditions for a second shock that compresses the material
again reaching densities that are much higher than can probed with
single shocks. If one starts from the high compression point on the
principal Hugoniot curve, one can reach densities of 65
g$\,$cm$^{-3}$. However, the compression ratio is typically not as
high because the strength of the interaction effects increases and
this lowers the compression ratio. For the secondary shock Hugoniot
curves that we show in Fig.~\ref{fig:DoubleShock}, the maximum compression
ratio varied between 4.51 and 3.97 while the maximum compression ratio
of the principlal Hugoniot curve was 4.66.

Fig.~\ref{fig:DoubleShock} also compares our secondary Hugoniot curves
with our isentropes and isotherms. For weak second shocks, the secondary
Hugoniot curves and isentropes almost coincide, which implies that the
second shock produces very little irreversible heat. As the
strength of the second shock increases, more and more irreversible
heat is generated.


\begin{figure}
    \centering
    \includegraphics[width=\columnwidth]{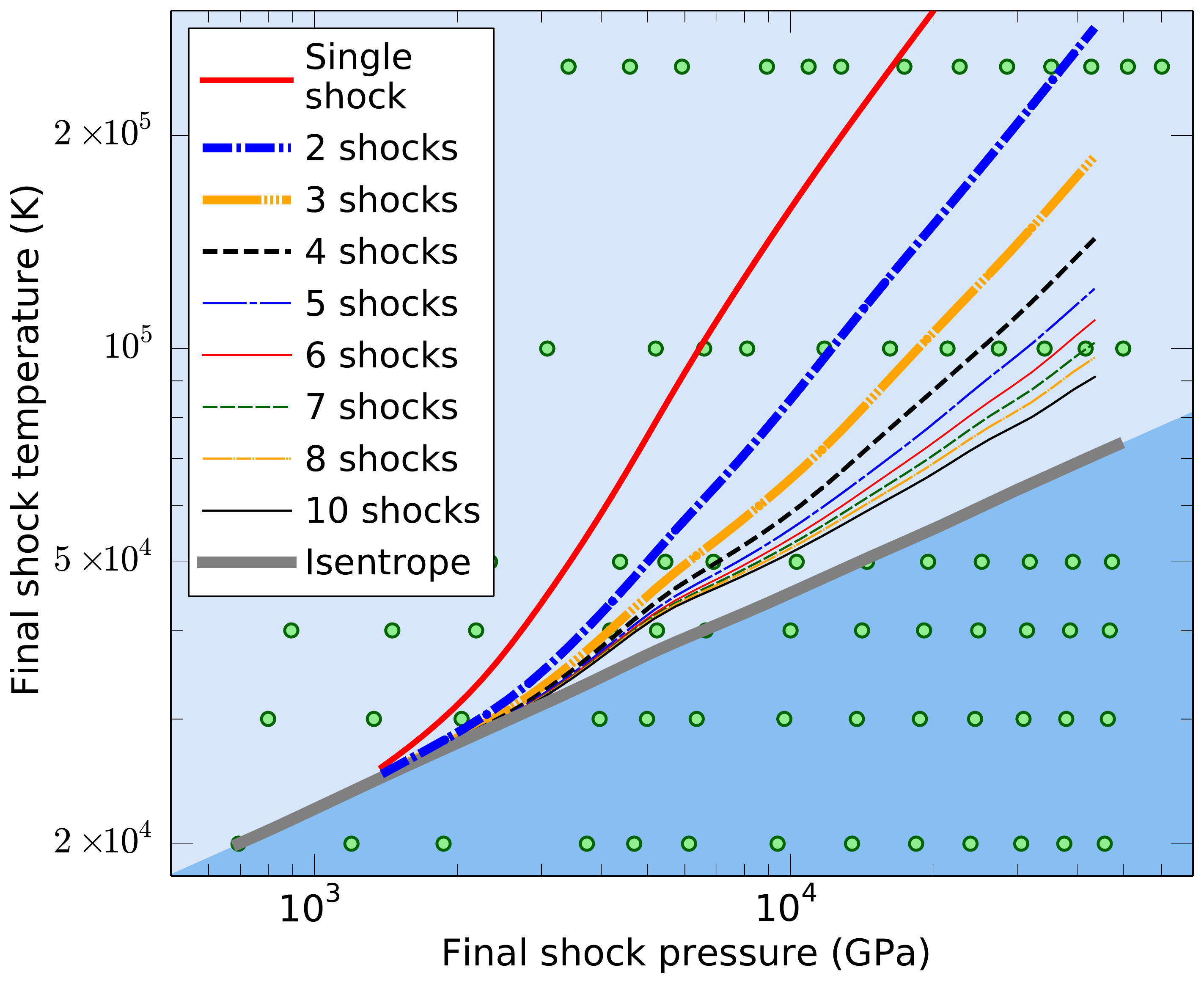}
    \caption{The amount of shock heating can be reduced if multiple shocks are used to compress a material rather than just one. Here we plot the final shock temperature as function of the final shock pressure for different numbers of shocks. The more shocks are employed, the closer the results are to an isentrope (thick grey line). 20$\,$000$\,$K and $\rho=2\rho_0$ were chosen as initial conditions. The circles represent the DFT-MD simulation points that we used in this analysis.}
    \label{fig:MultiShock}
\end{figure}

In Fig.~\ref{fig:MultiShock}, we compare an isentrope with various
multi-shock Hugoniot curves in order to determine how closely one can
trace an isentrope by breaking up a single shock into
multiple smaller shocks. All curves start from 20$\,$000$\,$K, twice the ambient density, and 1386 GPa. The isentrope~\cite{MH08b} was
derived from our EOS table using $\frac{dT}{dV}|_S = -T
{\frac{dP}{dT}|_V}/{ \frac{dE}{dT}|_V}$. For weak shocks, the Hugoniot
curve does not deviate much from an isentrope. For strong shocks, a
substantial amount of shock heating occurs. The resulting single-shock
Hugoniot curves are thus much hotter than an isentrope when both are
compared for the same pressure. The discrepancy depends significantly
on pressure, which reflects the fact that the final shock density
cannot exceed 4.7-fold the initial density (Fig.~\ref{fig:HugoniotComp}). To reach a large
final shock pressure, a substantial pressure contribution must come
from the thermal pressure, which requires high temperatures.

One can, however, get arbitrarily close to isentropic compression by
breaking up a single shock experiments into multiple weaker
shocks. The purpose of Fig.~\ref{fig:MultiShock} is to assess quantitatively how well
this method works for shocks in MgO and to determine how much shock
heating still occurs if the shock is broken up into $N=2$--$10$ steps. In
our multi-shock calculations, we successively solve Eq.~\ref{hug} to
connect the intermediate shock states. In order to obtain the lowest
possible shock temperature for a given number of shocks, we keep the
final shock pressure fixed while we carefully adjust the temperatures
of the intermediate shocks until we determined the global minimum of
the final shock temperature with sufficient accuracy.

As expected, all the resulting multi-shock Hugoniot curves converge to
an isentrope for weak shocks. For strong shocks such as
$P_{\rm final}/P_{\rm initial} \approx 31$, we find the temperature of single-shock
temperature is 8.36 times higher than the corresponding temperature on
the isentrope. If this shock is broken up into two, well-chosen
smaller shocks, the final shock temperature be reduced to 4.00 times the
value on the isentrope. If three, four, five, or even ten shocks are employed
the final shock temperature can, respectively, be reduced to 2.64,
2.02, 1.72, or 1.29 times the isentropic value. This are substantial
reductions compared to the single-shock temperatures.


\section{Conclusion}

By combining results from PIMC and DFT-MD simulations, we provided a consistent EOS table for MgO in the regime of warm, dense matter (see supplemental material), which includes conditions where the K- and L-shell electrons of both nuclei are ionized. The ionization of the L-shell only introduces a small increase in compression along the shock Hugoniot curve. The K shell ionization of both nuclei, however, introduces a broad temperature interval where shock compression exceeds values of 4.5. The maximum compression ratio of 4.66 is reached for conditions of 6$\,$378$\,$000$\,$K and 420$\,$900$\,$GPa where the K-shell ionization of the Mg nuclei dominates. By analyzing the electronic density of states in simulations at lower temperature we demonstrate the L-shells of Mg and O are strongly hybridized.

From the present work it is clear that mixing magnesium and oxygen under extreme conditions has non-ideal effects on the electronic structure and thus the ionization processes. It will make it challenging for other theoretical methods to approximate the properties of MgO by combining EOS tables of Mg and O without fully taking into account their interaction effects. But it also demonstrates that elements with a relatively close atomic number may be strongly influenced by one another. This could have some significance for ICF experiments and for stellar interior opacities where the radiative properties and thus the ionization stage is of essence.
\\

\begin{acknowledgments}
  This work was in part supported by the National Science
  Foundation-Department of Energy (DOE) partnership for plasma science
  and engineering (grant DE-SC0016248), by the DOE-National Nuclear
  Security Administration (grant DE-NA0003842), and the University of
  California Laboratory Fees Research Program (grant
  LFR-17-449059). F.S. was in part supported by the European Union through a Marie Sk\l odowska-Curie action (grant 750901). F.G.-C. acknowledges support from the CONICYT
  Postdoctoral fellowship (grant 74160058). Computational support was
  provided by the Blue Waters sustained-petascale computing project
  (NSF ACI 1640776) and the National Energy Research Scientific
  Computing Center.
\end{acknowledgments}


\bibliography{bibliography}

\end{document}